\documentclass[a4paper,10pt, twocolumn]{article}
\usepackage[T1]{fontenc} % if needed
\usepackage[utf8]{inputenc}
\usepackage{authblk}
\usepackage{blindtext}
\usepackage{abstract}
\usepackage{flushend} % to balance the 2 columns on the last page
\usepackage{graphicx}
\usepackage{xcolor}
\usepackage{pgf,pgfarrows,pgfnodes}
\usepackage{amsmath}
\usepackage{amssymb} 
\usepackage[section]{placeins} % This keeps floats inside their own section
\usepackage{upgreek} % Need this because particle names are labels and should NOT be italic
\title{Probing antigravitational effects\\through CP violation on the Moon}
\author[3]{G.~M.~Piacentino\thanks{giovanni.piacentino@unimol.it}}
\author[1,2]{A.~Gioiosa\thanks{antonio.gioiosa@gmail.com}}

\author[4]{A.~Palladino \thanks{palladin@bu.edu}}
\author[3]{V.~Testa \thanks{vincenzo.testa@oa-roma.inaf.it}}
\author[5]{G.~Venanzoni\thanks{graziano.venanzoni@lnf.infn.it}}
\affil[1]{Universit\`a degli Studi del Molise, Campobasso, Italy}
\affil[2]{INFN, Sezione di Roma2, Roma, Italy}
\affil[3]{INAF, Sezione di Monte Porzio Catone}

\affil[4]{Boston University, Boston, USA}
\affil[5]{INFN, Sezione di Pisa, Pisa, Italy}
\newcommand{\Klong}{\ensuremath{K_{\mathrm{L}}}}
\newcommand{\Kshort}{\ensuremath{K_{\mathrm{S}}}}
\newcommand{\Kshortlong}{\ensuremath{K_{\mathrm{S,L}}}}
\newcommand{\pion}{\ensuremath{\pi}}

\newcommand{\muon}{\ensuremath{\mu}}

\newcommand{\neutrino}{\ensuremath{\nu}}

\begin{document}
%\maketitle
%\flushbottom
\twocolumn[
 \begin{@twocolumnfalse}
 \maketitle
  \begin{abstract}
The environment on the Moon has numerous features that make it interesting not only for the study of astrophysical phenomena, but also elementary particle physics. In fact, vacuum conditions, low gravity, and exposure to a relatively intense irradiation of cosmic protons covering a large energy spectrum, make the lunar environment attractive for a wide range of particle physics experiments otherwise unworkable on Earth. We suggest one such experiment measuring the difference between the amount of CP violation as measured on the surface of the Earth and on the surface of the Moon, which could indicate quantum gravitational effects.
 \end{abstract}
\end{@twocolumnfalse}
]
\saythanks %this adds the email addresses at the bottom of the first page

\FloatBarrier
%<Orlando.Luongo@lnf.infn.it>
\section{Introduction}
The response of antimatter to a gravitational field has not yet been measured in any experimental test. A relatively large number of experiments on the gravitational interaction of anti-matter have been proposed and even started, e.g., ASACUSA~\cite{antihydrogen11, antihydrogen12, antihydrogen13, antihydrogen14, antihydrogen15}, ATHENA~\cite{antihydrogen2, antihydrogen8}, AEGIS~\cite{AEGIS-detector}, ALPHA~\cite{ALPHA-collaboration}, ATRAP~\cite{ATRAP-collaboration}, GBAR~\cite{GBAR-collaboration} and MAGE~\cite{MAGE-collaboration} to name only a few. Most of them are related with direct measurements of the interaction of anti-protons and of anti-nuclei and had, or have to, face technical difficulties due to copious electromagnetic noise and the difficulties of producing and confining anti-matter. MAGE deals with muonium and it is particullary interesting as the first experiment concerning the gravitational interaction of leptons proposed till now. Most of the ongoing experiments are still in the preliminary step of attempting to confine the anti-matter in order to perform their measurements \cite{antihydrogen1}..\cite{antihydrogen41}. ALPHA obtained antihydrogen confinement as long as 1000 seconds \cite{antihydrogen42}. This paper is focused on a different aspect of the problem; 
we are looking for a variation in CP~violation as a function of gravitational field intensity. Two experiments, CPLEAR~\cite{cplear-1999} and KLOE~\cite{mambriani_trentadue-2000}, dealt with CP violation in the neutral kaon system, both looking for a time modulation of CP violation due to tidal contributions of the Moon, Sun, and galaxy to the gravitational field involved in kaon decays. The results were not incompatible but too weak to see any correlation. 

In this paper, we propose a new approach to the problem, motivated by recent scientific interest in constructing a research village on the moon~\cite{taylor-1985,crawford-2012,mckay-2013}. In fact, the European Space Agency's (ESA) Director General recently suggested the creation of a human-robotic lunar outpost as a logical next step in human space exploration~\cite{woerner-2016, crawford-2017}.
Our approach takes advantage of the large difference between the gravitational field on the surface of the Earth and that on the Moon. On the Moon, the gravitational acceleration
$g_{\mathrm{Moon}} = 1.622$~m/s$^2$ is only $16.54\% $ of the corresponding acceleration on Earth. Such a difference should be extremely useful to our investigation on the contribution of gravity to the mixing in the neutral kaon system.

\section{Antimatter and Gravitation}
Any kind of relativistic and quantum theory of interactions, in the limit of perfectly static interactions, from the point of view of a Galileian inertial frame of reference, must converge to the classical expressions for the electromagnetic and gravitational fields,
\begin{equation}
F_{q,q^\prime} = \frac{1}{4\pi\varepsilon_0}\frac{qq^\prime}{r^{2}}\widehat{U}_{q,q^\prime}
\end{equation}
where $\varepsilon_{0} = 8.85 \times 10^{12}$~$\frac{\mathrm{C}^2}{\mathrm{N} \cdot \mathrm{m}^2}$, and
\begin{equation}
F_{m,m^\prime} = \frac{1}{4\pi g_0}\frac{mm^\prime}{r^{2}}\widehat{U}_{m,m^\prime}
\end{equation}
where $g_0 = -1.1935\times 10^{9}\frac{\mathrm{kg}^2}{\mathrm{N}\cdot\mathrm{m}^3}$. 

The different sign of the two constants, $e_0$ and $g_0$, means that similar charges repel for the electromagnetic interaction but attract for the gravitational interaction. This means that there is an electromagnetic screening effect, i.e., a bound state of electric charges seems to have less charge than the sum of the individual charges, while on the contrary a bound state of massive particles seems to have more mass from the point of view of an external observer (the mass of the proton is larger than the sum of those  of the valence quarks and this is true also in a gravitational system ~\cite{brillouin}). 

In the Newtonian formulation the definition of inertial mass is related to the definition of force and acceleration. Therefore this implies a dependence on the definition of time. Negative mass should be possible  in case of negative time. Particles with intrinsic reversed time evolution are  possible in the framework of the General Relativityi ~\cite{Ripalda} so that  it is possible to consider a repulsion between particle and antiparticle mediated by gravitation~\cite{hajdukovic-2012}. The definition of antimatter in quantum field theories was provided by Feynman and Stueckelberg~\cite{feynman-1987, stueckelberg-1942}. It implies that matter-antimatter transformations are equivalent to charge conjugation and time reversal in a flat space-time. When General Relativity (GR) is considered, space-time is not in general flat, however the GR equations remain symmetric in the time coordinate even if amtimatter is not explicitly considered in GR. Several theories that allow repulsion between matter and antimatter have been proposed~\cite{benoit-levy_chardin_2012, chardin-1993, villata-2011}, ~\cite{Ripalda} and ~\cite{hajdukovic-2012}. 
In fact, this hypothesis may overcome many of the open problems in astrophysics and cosmology: 
\begin{enumerate}
\item[\emph{i.}] In the observed universe, matter prevails over antimatter (Matter-Antimatter Asymmetry); % even if both are always created together; 
\item[\emph{ii.}] The Cosmic Microwave Background (CMB) is neither anisotropic nor inhomogeneous enough to be compatible with the Big Bang model without the introduction of an unknown interaction (Inflation);
\item[\emph{iii.}] Given the attractive gravitational interaction, we expect a slowing of the the expansion of the universe. On the contrary the expansion  seems to accelerate (Dark Energy);
\item[\emph{iv.}] The gravitational field of galaxies, clusters, and even our own solar system seems much stronger than expected due to visible matter (Dark Matter). 
\end{enumerate}

%\Most of them could be addressed by the same hypothesis because:
%\begin{enumerate}
%\item[\emph{i.}] An equal mix of matter and antimatter will give rise to a net repulsive force \textcolor{red}{(This needs to be explained, or provide a reference)}, this could generate inflation;
%\item[\emph{ii.}] Antigravity generated CP~violation could explain ``missing'' antimatter;
%\item[\emph{iii.}] Antigravity could also provide a mechanism for the acceleration of the universe (Einstein's constant / dark energy). %\textcolor{red}{(I think it would be better to explain these 3 points in more detail. About 3 sentences for each item.)}
%\end{enumerate}
Most of them could be addressed by the same hypothesis.
In fact, the net force in a universe with matter-antimatter symmetry should be repulsive because in such a case the forces do not necessarily cancel out. In an ionic solid, a very similar physical system, the same number of positive and negative charges are present, but the overall electrostatic force on the crystal is attractive (compensated by the Fermi exclusion principle). If we now visualize each positive ion as a particle, and each negative ion as an antiparticle, and replace the electrostatic interaction by the gravitational potential, we obtain from the Madelung-like model:
\begin{equation}
U_g = \frac{1}{2}N\alpha \frac{m^2}{R},
\end{equation}
where $U_g$ is the total gravitational energy, $N$ is the total number of particles with mass $m$, $R$ the separation between nearest antimatter, and $\alpha$ is the Madelung constant. For simplicity, and to keep the analogy with a crystal, $m$ and $R$ are assumed to be the same for the entire matter-antimatter mixture. The Madelung constant has values between 1.6 and 1.8 for most crystal structures. The overall force on the universe ($\mathrm{d}U_g/\mathrm{d}R$) is repulsive~\cite{Ripalda}.
Therefore, gravitational repulsion between matter and antimatter may have dramatic consequences in astrophysics and cosmology.

Several authors~\cite{benoit-levy_chardin_2012, chardin-1993, villata-2011} have studied possible consequences of anomalous antimatter gravity if we live in a symmetric universe, i.e., a universe with the same amount of matter and antimatter, with antimatter somehow hidden (for instance in the vacuum).
%\textcolor{red}{(What are ``the voids''? I never heard this term.)}).
According to a radically different direction of research~\cite{hajdukovic-2012}, we live in an asymmetric universe (i.e., matter dominates antimatter) but the gravitational properties of antimatter (negative gravitational mass) may have a major impact because of the quantum vacuum fluctuations interpreted as virtual gravitational dipoles with the potential to explain the nature of what we call dark matter. According to this model of the universe, the only matter-energy content of the universe is the Standard Model matter (i.e., matter composed of quarks and leptons interacting through the exchange of gauge bosons). Thus, the phenomena usually attributed to hypothetical dark matter may be considered as a consequence of the local variation of the gravitational polarization due to the baryonic matter immersed in the quantum vacuum.

\section{Antigravity and CP~Violation}
In 1961, Good~\cite{good-1961} calculated, using absolute potentials, that a repulsive gravitational interaction of antimatter should introduce a regeneration of kaons, thus resulting in an anomalously large level of CP~violation. At the time of his paper, CP~violation in the neutral kaon system was unknown and the argument appeared strong and elegant, but in light of the measurement of CP~violation~\cite{chronin_fitch-1964}, Good's paper has instead turned out to be a strong argument in favor of antigravity. 
Chardin reformulated Good's argument in terms of relative potentials and showed that the gravitational field on the surface of the Earth is of the right intensity, i.e., of the required order of magnitude to cause CP~violation. In particular, the mixing time of the kaon,

\begin{equation}
\Delta \tau = 5.9 \times 10^{10} \,\, \mathrm{s} \,\, \simeq 6\tau_{k_{s}}
\end{equation}
is long enough for the gravitational field of the Earth to attract the matter and repel the antimatter components of the $\mathrm{K} $ meson to induce a separation,
\begin{equation}
\Delta \zeta = g \Delta \tau^{2} \,\,,
\end{equation}
between them inducing regeneration, thus providing a mechanism for indirect CP~violation. The amount of such a contribution should be roughly the same order of magnitude as the spatial sepration divided by the Compton wavelength:
\begin{equation}
\label{eq:chi}
\begin{gathered}
\chi=\dfrac{\Delta \zeta}{\Delta L_{\mathrm{K}}}
=\dfrac{g \tau_{k_{l}}}{\dfrac{\hbar}{m_{\mathrm{K}}c}}
=\Omega\dfrac{g\dfrac{\pi^2\hbar^2}{\Delta m^2c^4}}{\dfrac{\hbar}{m_{\mathrm{K}}c}}\\ 
\chi=\Omega\dfrac{\pi^2\hbar gm_{\mathrm{K}}}{\Delta m^2c^3}=\Omega\times0.88\times10^{-3} \,\,,
\end{gathered}
\end{equation}
Here $\Omega$,
is a constant of the order of the unity representing the fraction of the decay time sensitive to gravitation.
So $\chi$
 happens to be of the same order of magnitude as the CP~violation parameter, $\varepsilon$, as measured on Earth's surface. 
If we calculate (\ref{eq:chi}) given the gravitational strength on the Moon's surface we would expect the measured $\varepsilon$ to be $80\%$ smaller than 
the $\varepsilon$ measured on Earth's surface, assuming gravitational repulsion beween matter and antimatter. 

\section{Moon Surface Experiment}
As mentioned above, we expect only $16\%$ of the CP~violation induced by gravity in an experiment performed in the gravitational field on the Moon. Since
\begin{equation}
\label{eq:R}
R = \frac{\Gamma(\mathrm{K}_{\mathrm{L}}\to \uppi^+\uppi^-)}{\Gamma(\mathrm{K}_{\mathrm{L}}\to \uppi^+\uppi^-\uppi^0)}
\end{equation}
\noindent is quadratic in $\varepsilon$, we expect a very large difference in the value of $R$ when measured on the surface of the Moon. In order to produce the $\mathrm{K}_{\mathrm{L}} $ on the Moon we plan to take advantage of the flux of cosmic protons continuously hitting the Moon's surface.
A direct measurement of the flux of protons on the lunar surface has not yet been made, but Ackermann et al.~\cite{ackerman-2016} fit the data of the gamma albedo from the Moon surface due to the incoming proton flux finding it to be equal, within a $10\%$ uncertainty, to the proton flux measured by AMS-02~\cite{ams-detector} and PAMELA~\cite{pamela-2017}. In the following, we considered the proton flux as determined by AMS-02.

In~\cite{piacentino-2016} and~\cite{piacentino-2017}, we considered an experimental apparatus with cosmic protons incident upon a cylindrical target followed by a detector region consisting of a cylindrical tracking volume (1~meter diameter, 1~meter deep) in a Low Earth Orbit (LEO). On a satellite in LEO, the effect due to gravity on $R$ is $\sim20\%$. If instead we place the apparatus on the surface of the Moon, we expect a significant improvement in the results, with an effect on $R$ of $\sim85\%$.

%5~meters in diameter and 5~meters high. The last 3~meters of the detector volume consist of a solenoidal magnetic field of $0.7$~T in order to distinguish charged particles. The end of the detector consists of an electromagnetic calorimenter 15~radiation lengths deep.
We performed Geant4 simulations using the angular and energy spectrum of the incident cosmic protons as measured by AMS-02 spectrometer. We simulated incident protons with $\theta_\mathrm{max} = 45^\circ $ over a target surface corresponding to a $\pi/4$ solid-angle acceptance. We used the same apparatus as in~\cite{piacentino-2016} and~\cite{piacentino-2017}, with the exception of the target. There, we used a cylindrical Tungsten target (1~m diameter, 9~cm deep), while here we considered an active target consisting of alternating layers of Tungsten and scintillating crystals (Stolzite, PbWO$_4$) for a total depth of 18~cm), to be read with photodiodes.

% Figures~\ref{fig:MaterialStudy} and~\ref{fig:BestThickness} show results from an optimization study of target depth. 
% In Figure~\ref{fig:MaterialStudy} we see that even though more $\mathrm{K}_{\mathrm{L}} $ are produced for thicker targets, the probability that they exit the thicker targets is reduced due to regeneration and nuclear interactions. Figure~\ref{fig:BestThickness} shows the number of $\mathrm{K}_{\mathrm{L}}$ decays downstream of the target for various target materials and thicknesses. 

We studied the production of $\mathrm{K}_{\mathrm{L}}$ that would decay inside the volume of a detector and their distribution on the detector cross section. The particles decaying before $z = 18$~cm decay inside the PbWO$_4$ target and are lost, however those decaying inside a downstream cylindrical tracking region could potentially be reconstructed. At this stage in evaluating the the feasibility of performing the experiment, we made no attempt to fully reconstruct the $\mathrm{K}_{\mathrm{L}}$ decay products inside the detector. We just took as a crude approximation the reconstruction efficiency as equal to~1 inside the fiducial volume of the tracking region. With protons incident on the target distributed in angle and energy as cosmic protons, we estimate the number of $\mathrm{K}_{\mathrm{L}}$ decays per year inside two possible tracking volumes, listed in Table~\ref{tab:NKL}, with and without additional kinematical cut on the axial momentum at the decay vertex. Table~\ref{tab:sim_results} shows the length of time it would take to record sufficient $\mathrm{K}_{\mathrm{L}}$ decays to provide 3$\sigma$ and 5$\sigma$ measurements of $R$.

% \begin{table*}[t]
% \centering
% \caption{\label{tab:table1}Number of $\mathrm{K}_{\mathrm{L}}$ that decay within various tracking region sizes. This Monte Carlo simulation consisted of a 1~m diamter, 
% 18~cm thick PbWO$_4$ target and an angular acceptance of incoming protons within 45~degrees. The flux and energy distribution of incident protons were generated according to the AMS-01 results.}
% \begin{tabular}{l*{6}{c}r}
% Detector size (diameter $\times$ depth) & 1m $\times$ 1m & 1.5m $\times$ 2m & 5m $\times$ 5m \\
% \hline
% $\mathrm{K}_{\mathrm{L}}$ decays per year & $1.09 \times 10^{6}$ & $2.11 \times 10^{6}$ & $6.68 \times 10^{6}$ \\
% $\mathrm{K}_{\mathrm{L}}\to\uppi^+\uppi^-\uppi^0$ decays per year & $1.42 \times 10^{5}$ & $2.74 \times 10^{5}$ & $8.67 \times 10^{5}$ \\
% \end{tabular}
% \end{table*}

\begin{table}[h]
\tiny
\centering
\begin{tabular}{l*{6}{c}r}\hline
Volume, & ($r<50$~cm) & ($r<50$~cm, & ($r<100$~cm, \\
kinematics & {} & $p_z<0.5$~GeV/c) & $p_z<0.5$~GeV/c) \\
\hline
$\frac{N\left(\Klong\text{decays}\right)}{\text{year}}$ & $3.54 \times 10^{6}$ & $1.50 \times 10^{6}$ & $3.36 \times 10^{6}$ \\
$\frac{N\left(\Kshort\text{decays}\right)}{N\left(\Klong\text{decays}\right)}$ & $3.03 \times 10^{-1}$ & $4.15 \times 10^{-5}$ & {} \\
\hline
\end{tabular}
\caption{\label{tab:NKL}Number of $\mathrm{K}_{\mathrm{L}}$ that decay within various tracking region volumes
50~cm downstream of the target ($50<z<150$~cm, $r<50$~cm or $r<100$~cm), with and without a kinematical cut on 
the axial momentum at the $\Kshortlong$ decay vertex.}
\end{table}

% \begin{table}[h]
% \centering
% \caption{\label{tab:sim_results}Critical parameters necessary for $3\sigma$ and $5\sigma$ measurements of a 10\% change in the level 
% of CP~violation (20\% change in $R$) along with the values obtained from our Monte Carlo simulation. The results take into account a 
% basic geometrical event selection of $\Kshortlong$ decay vertices within a 1~m $\times$ 1~m cylindrical 
% tracking volume 50~cm downstream of the target ($50<z<150$~cm, $r<50$~cm).
% and axial momentum at the $\Kshortlong$ decay vertex of $p_{z}<0.5$~GeV/c.
% These values assume a 100\% detection efficiency, 2\%~(4\%) statistical and 2\%~(4\%) systematic fractional
% uncertainties for 5$\sigma$~(3$\sigma$).}
% {\small
% \begin{tabular}{c|c|c|c|c}
% \hline \hline
% {} & \multicolumn{2}{ c| }{ Requirement } & \multicolumn{2}{ c }{ Simulation result } \T\B \\
% {} & 3$\sigma$ & 5$\sigma$ & 3$\sigma$ & 5$\sigma$ \\
% \hline 
% $ N\left( \Klong\text{ decays} \right)$ & $>3 \times 10^5$ & $>12.5\!\times10^5$ & 73 days & 304 days \T\B \\ %\hline %\noalign{\vskip 1mm} 
% $\frac{ N\left( \Kshort\text{ decays} \right) }{ N\left( \Klong\text{ decays} \right) }$ & $<1 \times 10^{-4}$ & $<5.7\!\times10^{-5}$ & \multicolumn{2}{ c }{ $4.1 \times 10^{-5}$ } \T\B \\ %\hline %\noalign{\vskip 1mm} 
% $\frac{\delta N\left( \Klong\to\pion\muon\neutrino \right) }{ N\left( \Klong\to\pion\pion \right) }$ & $<4 \times 10^{-2}$ & $<2\!\times10^{-2}$ & \multicolumn{2}{ c }{ kinematical cuts } \T\B \\ \hline %\noalign{\vskip 1mm} 
% \hline
% \end{tabular}
% }
% \end{table}
\begin{table}[h]
\centering
\tiny
\begin{tabular}{c|c|c|c|c}
\hline
{} & \multicolumn{2}{ c| }{ Requirement } & \multicolumn{2}{ c }{ Simulation result } \\
{} & 3$\sigma$ & 5$\sigma$ & 3$\sigma$ & 5$\sigma$\\
\hline 
$ N\left( \Klong\text{ decays} \right)$ & $>2.5 \times 10^4$ & $>7\times10^4$ & 6 days & 17 days\\ %\hline %\noalign{\vskip 1mm} 
$\frac{N\left( \Kshort\text{ decays} \right) }{ N\left( \Klong\text{ decays} \right) }$ & $<1 \times 10^{-4}$ & $<5.7\!\times10^{-5}$ & \multicolumn{2}{ c }{ $4.15 \times 10^{-5}$ } \\ %\hline %\noalign{\vskip 1mm} 
$\frac{\delta N\left( \Klong\to\pion\muon\neutrino \right) }{ N\left( \Klong\to\pion\pion \right) }$ & $<4 \times 10^{-2}$ & $<2\!\times10^{-2}$ & \multicolumn{2}{ c }{ kinematical cuts } \\ \hline %\noalign{\vskip 1mm} 
\end{tabular}
\caption{\label{tab:sim_results}Critical parameters necessary for $3\sigma$ and $5\sigma$ measurements of a gravitational modulation in the level of CP~violation (85\% change in $R$) along with the values obtained from our Monte Carlo simulation. The results take into account a 
basic geometrical event selection of $\Kshortlong$ decay vertices within a 1~m $\times$ 1~m cylindrical tracking volume 50~cm downstream of the target ($50<z<150$~cm, $r<50$~cm),
and axial momentum at the $\Kshortlong$ decay vertex of $p_{z}<0.5$~GeV/c.
These values assume a 100\% detection efficiency, 2\%~(4\%) statistical and 2\%~(4\%) systematic fractional uncertainties for 5$\sigma$~(3$\sigma$).}
\end{table}

\section{Conclusions}
By constructing a detector consisting of a 1~m diameter, 18~cm thick active target and a 1~m diameter $\times$ 1~m deep tracking and particle identification system, and placing it on the surface of the Moon, we could perform a direct measurement of the ratio of the number of $\mathrm{K}_{\mathrm{L}}$ decaying to two charged pions to those decaying to three pions in a low-gravity environment. We estimate that it will take 6~days (17~days) to record sufficient $\mathrm{K}_{\mathrm{L}}$ decays for a 3$\sigma$ (5$\sigma$) measurement. Any difference between the amount of CP~violation on the surface of the Moon compared to the level of CP~violation on the surface of Earth would be an indication of a quantum gravitational effect.

% \begin{figure}[t]
% \begin{minipage}[t]{0.48\textwidth}
% \centering
% %\includegraphics[width=.98\textwidth,trim=0 0 0 25,clip]{./figures/material_study.pdf}
% \caption{\label{fig:MaterialStudy} Number of $\mathrm{K}_\mathrm{L}$ produced for a variety of target materials.
% }
% \end{minipage}
% \end{figure}

% \begin{figure}[t]
% \begin{minipage}[t]{0.48\textwidth}
% \centering
% %\includegraphics[width=.98\textwidth,trim=0 0 0 25,clip]{./figures/c0.pdf}
% \caption{\label{fig:BestThickness} Number of $\mathrm{K}_\mathrm{L}$ decays downstream of the target for various target materials and thicknesses.}
% \end{minipage}
% \end{figure}

% \begin{figure}[t]
% \begin{minipage}[t]{0.48\textwidth}
% \centering % \begin{center}/\end{center} takes some additional vertical space
% %\\includegraphics[width=.98\textwidth,trim=0 0 0 25,clip]{./figures/KL_decay_vertices.pdf}
% \caption{\label{fig:decayVertex} Position of $\mathrm{K}_\mathrm{L}$ decay vertices
% for a simulation consisting of a cylindrical PbWO$_4$ target (50~cm radius, 18~cm thick) indicated
% with the red box. The blue box outlines a cylindrical tracking region with 75~cm radius and 2~m depth, shown
% as a reference.}
% \end{minipage}
% \end{figure}

%\acknowledgments
%Write some acknoledgments here.

%\paragraph{Note added.} This is also a good position for notes added
%after the paper has been written.

% BIBLIOGRAPHY
% use BIBTEX if you want
%\bibliographystyle{JHEP}
%\bibliography{yourBIBfiles}

\end{document}